\documentclass[pra,floatfix,twocolumn,superscriptaddress]{revtex4-1}
\usepackage[final]{graphicx}
\usepackage{times,bbm,amsmath,amssymb}
\usepackage{epsfig,color}
\usepackage{hyperref}
\usepackage{float,siunitx}
\usepackage[caption = false]{subfig}
\usepackage[greek,english]{babel}
\usepackage{thumbpdf,enumerate}
\usepackage{booktabs}
\usepackage{sidecap}
\usepackage[scaled=.8]{couriers}
\usepackage{pstricks}
\usepackage{multirow}
\usepackage{placeins}
\usepackage{pst-grad}
\usepackage{epigraph}
\usepackage{longtable}
\usepackage{booktabs}
\usepackage{gensymb}

\newcommand{\Tr}{{\rm Tr}}

\def\Tr{\hbox{Tr}} 
\usepackage{pifont}

\newcommand{\ket}[1]{\vert#1\rangle}
\newcommand{\bra}[1]{\langle#1\vert}

\begin{document}

\title{Information-thermodynamics of Quantum Generalized Measurements}

\author{Luca Mancino}
\affiliation{Dipartimento di Scienze, Universit\`{a} degli Studi Roma Tre, Via della Vasca Navale 84, 00146, Rome, Italy}

\author{Marco Sbroscia}
\affiliation{Dipartimento di Scienze, Universit\`{a}  degli Studi Roma Tre, Via della Vasca Navale 84, 00146, Rome, Italy} 

\author{Emanuele Roccia}
\affiliation{Dipartimento di Scienze, Universit\`{a} degli Studi Roma Tre, Via della Vasca Navale 84, 00146, Rome, Italy}

\author{Ilaria Gianani}
\affiliation{Dipartimento di Scienze, Universit\`{a} degli Studi Roma Tre, Via della Vasca Navale 84, 00146, Rome, Italy}

\author{Fabrizia Somma}
\affiliation{Dipartimento di Scienze, Universit\`{a} degli Studi Roma Tre, Via della Vasca Navale 84, 00146, Rome, Italy}

\author{Paolo Mataloni}
\affiliation{Quantum Optics Group, Dipartimento di Fisica, Universit`a di Roma La Sapienza, Piazzale Aldo Moro 5, I-00185 Roma, Italy.}

\author{Mauro Paternostro}
\affiliation{Centre for Theoretical Atomic, Molecular and Optical Physics, School of Mathematics and Physics, Queen's University Belfast, Belfast BT7 1NN, United Kingdom}

\author{Marco Barbieri}
\affiliation{Dipartimento di Scienze, Universit\`{a} degli Studi Roma Tre, Via della Vasca Navale 84, 00146, Rome, Italy}

\begin{abstract}
Landauer's principle introduces a symmetry between computational and physical processes: erasure of information, a logically irreversible operation, must be underlain by an irreversible transformation dissipating energy. Monitoring micro- and nano-systems needs to enter into the energetic balance of their control; hence, finding the ultimate limits is instrumental to the development of future thermal machines operating at the quantum level. We report on the experimental investigation of a bound to the irreversible entropy associated to generalized quantum measurements on a quantum bit. We adopted a quantum photonics gate to implement a device interpolating from the weakly disturbing to the fully invasive and maximally informative regime. Our experiment prompted us to introduce a bound taking into account both the classical result of the measurement and the outcoming quantum state; unlike previous investigation, our new entropic bound is based uniquely on measurable quantities. Our results highlight what insights the information-theoretic approach can provide on building blocks of quantum information processors.
\end{abstract}

\maketitle

\textit{Introduction.} Manipulating the information content of a memory register has an associated thermodynamic cost: the erasure and copying of information can only be realized through physical operations whose implementation necessitates adequate resources \cite{Plenio01,Piechocinska00,Lloyd00,Bennett03,Maruyama09}. Landauer's principle epithomizes such correspondence by quantifying the minimum amount of work required to reset a memory register \cite{Landauer61,Landauer91} and thus entering the energetic balance of any physical process. 

Controlling the conversion of information to energy will be key to the implementation of quantum-limited micro- and nano-engines and, in general, to the development of quantum technologies \cite{QuantumManifesto}. In the light of Landauer's principle, the simple act of monitoring a system through a quantum measurement implies a thermodynamic cost whose quantification has been the focus of recent debate \cite{Sagawa09,Granger11,Jacobs12,Alonso15,Auffeves16}. 

Any quantum measurement introduces a disturbance to the state of the monitored system. This is a key issue in the control of the dynamics of the latter, which requires the optimization of the trade-off between the intrusivity of a measurement and the degree of information acquired through it. Weak measurements \cite{Aharonov88,Aharonov89} are valid tools to achieve such a compromise as experimentally demonstrated in the context of metrology~\cite{Hosten08,Dixon09,Brunner10,Zhang15}, communication \cite{Levenson93,Biggerstaff09,Gillett10}, and error correction \cite{Steane96,Blok14}. 

Here we investigate the thermodynamic cost of generalized measurements in a photonic architecture \cite{Pryde05,Ralph06}. Starting from the framework set in Ref.~\cite{Sagawa09}, we introduce a new figure for quantifying the efficiency of the measurement which is attainable in a real experiment. The flexibility of our experimental setup allows for the interpolation between weakly disturbing and strong measurements. Our results show the energetic equivalence of performing measurements of arbitrary invasiveness. At variance with this, a weak measurement impacts in a less substantial way on the coherence of the quantum state of the measured system. While extending Landauer-like arguments, our endeavors open up the exploitation of generalized measurements in information-to-energy conversion processes. 

\begin{figure}[t!]
\includegraphics[width=\columnwidth]{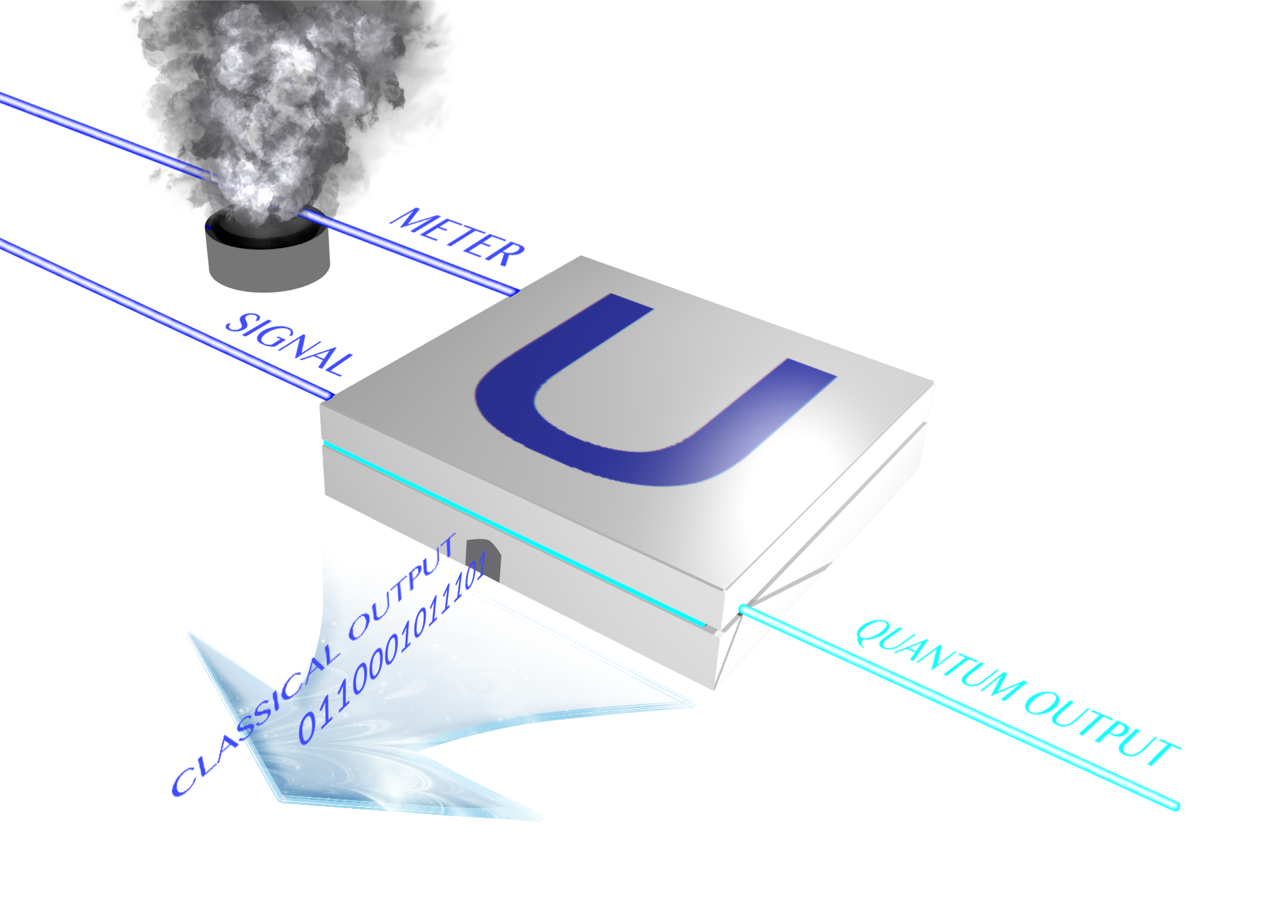}
\caption{Conceptual scheme of the protocol. After the meter is kept in contact with a thermal bath at fixed temperature T, the two systems interact through a unitary operator. It is possible to infer information on the signal by performing a measurement on the meter resulting in a classical and a quantum output.}
\label{Protocol}
\end{figure}

\textit{Results.} In Fig. \ref{Protocol}, we show the salient features of our protocol. A quantum system, initially in the quantum state $\rho_{\sigma}$, is sent to a non destructive measurement device. Information is extracted by looking at an ancillary meter system, which we assume has initially been put in contact with a thermal bath at temperature $T$. As a result, the meter is prepared in the canonical Gibbs state $\rho_{\mu}=e^{-\beta \mathbb{H}_{\mu}}/\mathbb{Z_{\mu}}$, where $\beta=1/k_B T$ (here $k_B$ is the Boltzmann constant), $\mathbb{H}_{\mu}$ is the free Hamiltonian of the meter, and $\mathbb{Z}_\mu={\rm Tr}[e^{-\beta \mathbb{H}_{\mu}}]$ is the partition function (see Methods).

The measurement relies on the interaction of the signal and the meter ruled by a unitary operation $U$. The coupling links the two systems in such a way that a standard measurement on the meter delivers information on the state of the signal. Such information is in the form of a \textit{classical} outcome $k$ occurring with a probability $p_k$. The amount of classical information is then quantified by the Shannon entropy $H(\lbrace p_k \rbrace)$, associated with the probabilities $p_k=\Tr[E_k \rho_\sigma]$, where we have introduced the set of measurement operators $E_k$ such that $\sum_kE_kE^\dag_k=\openone$. 
Remarkably, the Shannon entropy has a clear thermodynamic interpretation as formulated by Landauer's principle: an agent having the signal at their disposal might try to extract work from it by exploiting the information gathered through the measurement performed over the meter. The amount of work extractable from the measurement scheme being considered is then $k_B T(1-H(\lbrace p_k \rbrace))$~\cite{Maruyama05,Ciampini16}. 

We also have a {\it quantum} output, represented by the state of the signal $\rho_{\sigma}^{(k)}$ {\it after} the interaction, conditioned on the outcome $k$. A complete thermodynamic analysis ought to account the transformation of the signal system. A first analysis has been carried out in Ref.~\cite{Sagawa09}: here we go beyond the context and results set by such seminal work to address quantities that  are directly accessible to (and relevant for) the experimenter.

\begin{figure*}[t!]
\includegraphics[width=0.8\textwidth]{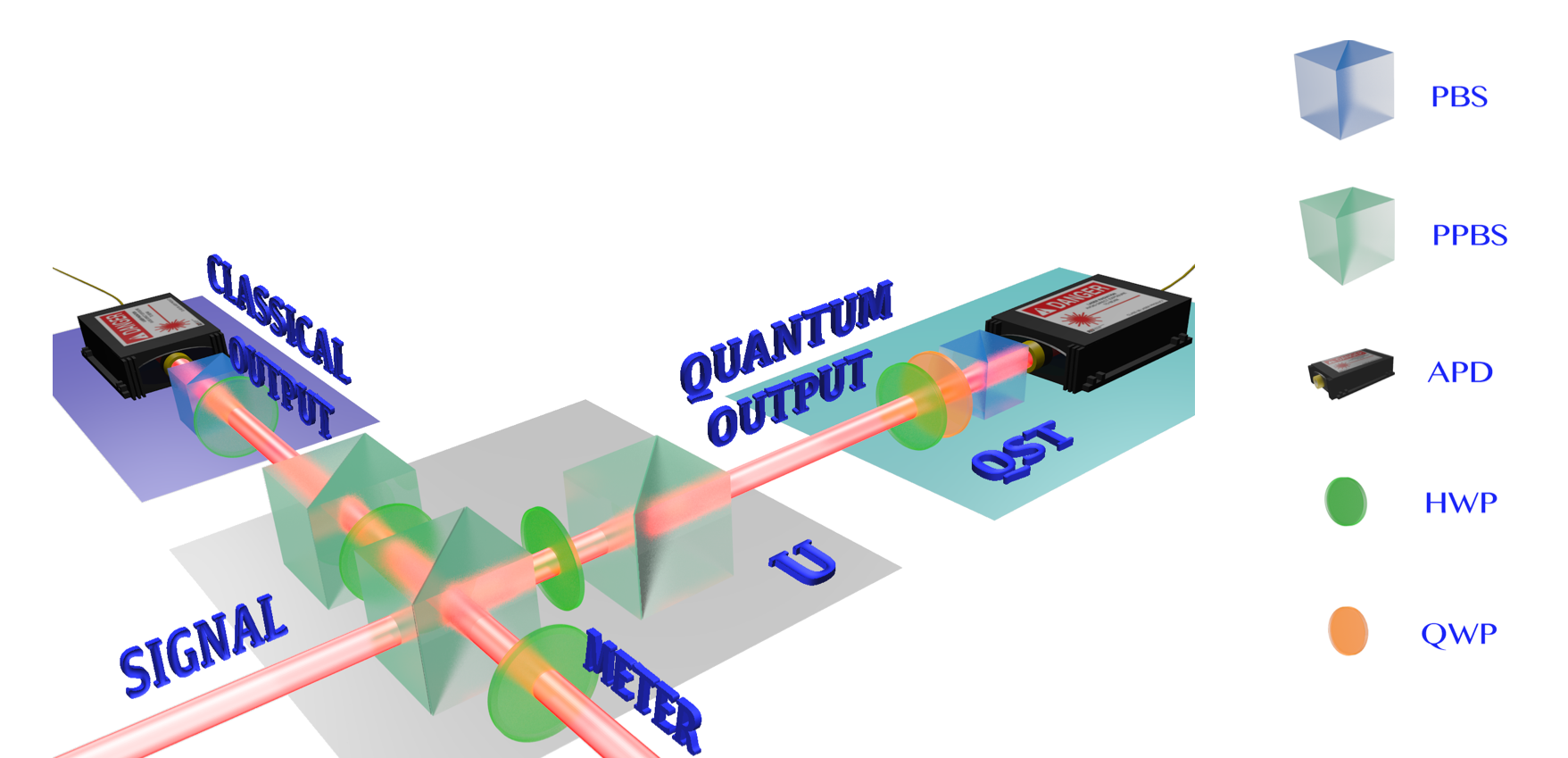}
\caption{Experimental setup. The signal and meter photons are generated via a Spontaneous Parametric Down Conversion (SPDC) process through a $\beta$ Barium-Borate crystal (Type I) pumped with a Continuous-Wave (CW) laser at 80 mW and wavelength 405 nm. We encode the logical states of each qubit in the horizontal and vertical polarization states $\vert H \rangle$ and $\vert V \rangle$ of each photon. The signal qubit is kept in the superposition state $\vert D \rangle=1/\sqrt{2} (\vert H \rangle + \vert V \rangle)$ while the meter starts in a $\vert H \rangle$ polarization state. The measurement strength can be adjusted with a rotation of the polarization of the meter performed via a Half Wave Plate (HWP). The two photons then enter in an entangling C-Sign gate realized with Partially Polarized Beam Splitters (PPBSs) and HWPs (see Methods). The classical output is then measured in the diagonal basis while on the quantum output a Quantum State Tomography (QST) is performed.}
\label{Setup}
\end{figure*}

The key figure of merit is the Groenewold-Ozawa (GO) or Quantum-Classical (QC) information~\cite{GO}
\begin{equation}
\label{balance}
\begin{aligned}
I&=S(\rho_{\sigma})+ H(\{ p_k \}) \\
&+ \sum_{k} \Tr\left[\left( \sqrt{E_{k}} \rho_{\sigma} \sqrt{E_{k}} \right) \ln \left( \sqrt{E_{k}} \rho_{\sigma} \sqrt{E_{k}} \right)\right].
\end{aligned}
\end{equation} 
This is the sum of three contributions: the von Neumann entropy of the initial signal state $S(\rho_{\sigma})$, the aforementioned Shannon entropy $H(\{ p_k \})$, and a term describing the residual entropy of the post-measurement states. The energetic and informational balance resulting from the process being considered is condensed in the expression of the ensemble-average work $W_\mu^{meas}$ needed for implementing the measurement itself. The balance can be cast into the form~\cite{Sagawa09}
\begin{equation}
\label{SU}
\beta (W_{\mu}^{meas}- \Delta F_{\mu})\geq   (I-H) 
\end{equation}
where $\Delta F_{\mu}$ is the average variation in the Helmholtz free energy of the meter in the overall process. The complete thermodynamic energetics can thus be understood in terms of an irreversible entropy (see Methods), embodied by the left-hand side of the inequality above that is dimensionally an entropy, generated by the process implemented by the measurement, whose lower bound is of a genuine information-theoretical nature and set at $I-H$. The implications of Eq.~\eqref{balance} have not been explored, nor the lower bound investigated experimentally.

\begin{figure}[b!]
\includegraphics[width=\columnwidth]{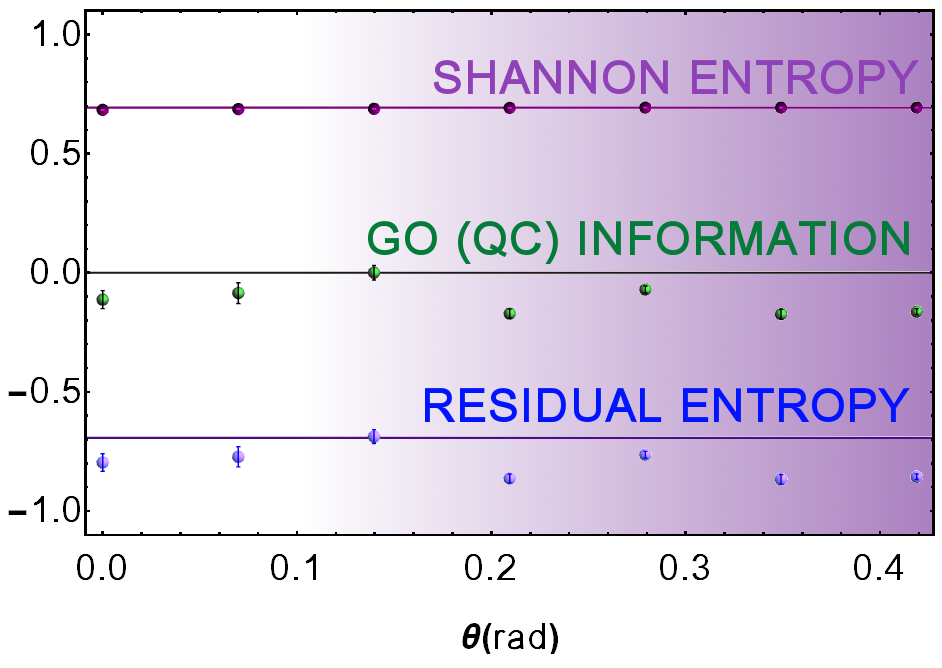}
\caption{Experimental behaviour at $T=0$. The experimental values of the Shannon entropy (purple), the GO (QC) Information (green), and the correlation term or residual entropy of the post-measurement states (blue) are displayed in function of the measurement strength $\theta$; the corresponding continuous lines represent the predictions for an ideal measurement. Vertical error bars take into account the Poissonian statistics of the measured counts and derived from a Monte Carlo (MC) routine.}
\label{Tug0}
\end{figure}

Here, we apply this framework to a two-photon experiment, with the aim of characterizing generalized measurements using information-thermodynamics.  For our investigation, we use the experimental setup depicted in Fig.~\ref{Setup}, where the polarization states photon pairs are used to encode the states of the signal and meter systems, assumed to be two-level system in the remainder of this work. The least (most) energetic state of each system is encoded in the horizontal $\vert H \rangle$ (vertical $\vert V \rangle$) polarization state of the respective photon. 

\begin{figure*}[t!]
\includegraphics[width=0.8\textwidth]{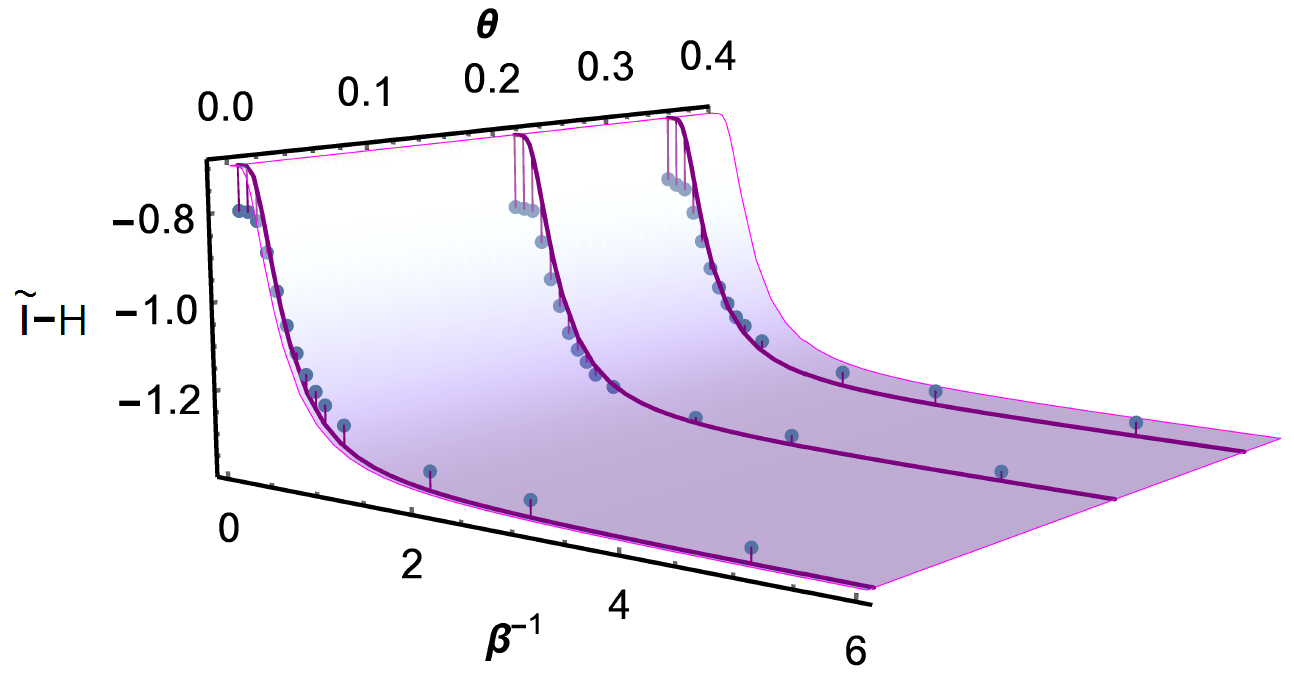}
\caption{Experimental behaviour at $T\neq0$. The figure shows the comparison between the theoretical prediction and the experimental values against the temperature of the meter ($\beta^{-1}$) and the measurement strength ($\theta$). Albeit the theoretical prediction does not depend on $\theta$, experimentally we notice that for low temperatures the impact of the setup imperfections is more pronunced in the strong regime as degree of entanglement between the signal and the meter increases. As the temperature rises, this effect becomes negligible. Vertical error bars are smaller than the point size.}
\label{Tdiv0}
\end{figure*}

Our experiment proceeds in four steps: i) the signal is initialized in a superposition of its levels $\vert D \rangle = \frac{1}{\sqrt{2}}\left( \vert H \rangle + \vert V \rangle \right)$ that is expected to deliver the highest entropy, while the meter is prepared in either the $\vert H \rangle$ or  $\vert V \rangle$ state; ii) the coupling is implemented by a Controlled-Sign (C-Sign) gate, which imparts a $\pi$-phase on the $\ket{V}_{\sigma} \ket{V}_{\mu}$ component only. Its strength is effectively controlled by rotating the state of the meter by means of a Half Wave Plate (HWP) before the gate, set at a rotation angle $\theta$. We can then create an entangled state of the two systems, so that a measurement on the meter will deliver information on the $H/V$ component of the signal without destroying it; iii) for different settings of $\theta$, we measure the probabilities $p_D$ and $p_A$ of a standard polarization measurement performed on the meter. This allows us to compute the Shannon entropy associated to each coupling: we can then explore different regimes, ranging from strong von Neumann measurements to weak observations that minimally disturb the input signal state; iv) we perform quantum state tomography of the signal photon conditioned on the outcome of the nondestructive measurement process. All relevant quantities are extracted from coincidence counts originating from a signal and a meter photon. This allows us to control the two out-coming states and use the results to evaluate the residual entropy of the post-measurement states.

First, we discuss the limiting case in which the meter is kept in the pure state $\ket{H}$. Our results are summarized in Fig.~\ref{Tug0}, where we show the Shannon entropy, the residual post-measurement entropy, and the GO (QC) information as functions of the angle $\theta$ corresponding to different measurement regimes. Clearly, $\theta=0$ corresponds to no coupling, while $\theta=\pi/8$ corresponds to the von Neumann projective regime. For all measurement strengths, the GO (QC) information ($I$) is expected to be equal to zero, given that the post-measurement entropy term reduces to the opposite of the Shannon entropy. We can develop an intuition for this result, considering that the energetics of the process Eq.~\eqref{SU} should be governed only by the statistics of the measurement, which is the only source of entropy. The experiment reveals that imperfections impact these expectations, by introducing an excess of von Neumann entropy on the post-measurement states of the signal, while the behaviour of the Shannon entropy remains close to the theoretical predictions. Remarkably, the strength of the measurement does not come into play when inspecting the balance equation.

More generally, the state of the meter is assumed to be in a canonical distribution at temperature T, due to its interaction with a thermal bath. We simulate this mixed state by adding, with suitable weights, the concidence counts relative to the least ($\vert H \rangle$) and the most ($\vert V \rangle$) energetic states of the meter, while keeping the signal prepared in $\vert D \rangle$. 

The GO (QC) information term is not directly accessible through the experiment. The reason is that the overall measurement apparatus actually performs distinct operations, depending on the preparation of the meter. The operator describing the post-measurement state of the signal when a meter photon prepared in $\vert H \rangle$ is injected is ${M_k \rho_{\sigma}^{k} {M_k}^\dagger}/p_k$ with $M_k^{\dagger} M_k = E_k$. Conversely, when a meter $\vert V \rangle$ is chosen, the state of the signal associated to the same output is now ${N_k \rho_{\sigma}^{k} {N_k}^\dagger}/p_k$, where $N_k=-i\sigma_y M_k$. However, we still have $N_k^{\dagger} N_k = E_k$. As the meter is kept in a mixed state, the measurement will output a mixture of the aforementioned states of the signal. The definition of the GO (QC) information in Eq.~\eqref{balance}, instead, considers only the case where $M_k=N_k=\sqrt{E_k}$, thus resulting in the discrepancy between the expected and observed values of such contribution. 

In order to go beyond such limitations and consider a figure of merit that is more faithful to the experimental setting being studied, we introduce a new (related) figure of merit defined as 
\begin{equation}
\label{Revisited}
\tilde{I} = S(\rho_{\sigma})+ H(\{ p_k \}) + \sum_{k} \Tr[\tilde{\rho}_{\sigma}^k \ln \tilde{\rho}_{\sigma}^k ],
\end{equation}
where $\tilde{\rho}_{\sigma}^k=1/2\mathbb{Z}_{\mu} \left[ e^{-\beta \epsilon_0} (M_k\rho_{\sigma} M_k^{\dagger}) +  e^{-\beta \epsilon_1} (N_k\rho_{\sigma} N_k^{\dagger}) \right]$ is the actual output state of the signal up to normalization. Such a state can be accessed experimentally by quantum state tomography. 

The results are summarized in Fig. \ref{Tdiv0}, which shows how the increase in the temperature of the meter, thus its mixedness, reflects in the dispersion of the correlation term $\tilde{I}-H$. The data confirm that this phenomenon is insensitive to the measurement strength: the degree of irreversibility generated by the implementation of the measurement addressed in our experiment is lower bounded by a quantity that is insensitive of the information gathered on the state of the signal through the meter. Thermodynamically, this implies that the minimum cost for the implementation of a measurement, as measured by the information-theoretical lower bound in Eq.~\eqref{SU} or the version proposed here involving $\tilde{I}$ is not linked to the back-action induced to the state of the probed system, but intrinsic in the act of measuring itself. 

On one hand, this calls for the research of bounds to the cost of measuring that are more sensitive to the explicit degree of back-action induced by the measuring step, much along the lines of recent attempts made in the context of the Landauer principle itself~\cite{Reeb,Goold}. On the other hand, $\tilde{I}$ accounts explicitly for the impossibility of an experimental apparatus such as ours to distinguish among the conditional states resulting from the recording of a given measurement outcome. Its introduction highlights the need for experiment-tailored quantities apt to quantify appropriately the energetic and information balance. 

{\it Conclusions.} We have investigated and characterized experimentally the minimum entropy cost necessary for the implementation of the measurement on the quantum state of an elementary system. Our experiment has been able to highlight the fundamental insensitivity  of such entropic bound to the invasiveness of the measurement itself. Our approach is based on the use of generalized quantum measurements, spanning from weak to strong projective ones. It demonstrates the viability of such an important tool for experimental investigations on the information-thermodynamics of fundamental quantum processes.

\bigskip

\textit{Methods: theoretical details.} The free Hamiltonian of the meter is assumed to be, without loss of generality, $\mathbb{H}_\mu=\epsilon_0 \ket{H}\bra{H} + \epsilon_1 \ket{V}\bra{V}$, with $\epsilon_0 (\epsilon_1)=0(1)$. The key step of our protocol is represented by the interaction of the signal and the meter via the unitary operator $\mathbb{U}=(\mathbb{I}\otimes\mathbb{R})\mathbb{N}$. Here, i) $\mathbb{R}$ is a rotation gate whose action allows to control the invasiveness of the measurement process \cite{Pryde05,Ralph06}; ii) $\mathbb{N}$ is the operator associated to the C-Sign gate.

When the meter is in state $\rho_{\mu}=\ket{H}\bra{H}$, the gate delivers a measurement operator acting like
\begin{equation}
M_k\rho_{\sigma} M_k^{\dagger}=\vert \psi_k \rangle \langle \psi_k \vert,
\end{equation}
where
~\cite{Ralph06}:
\begin{equation}
\begin{aligned}
\vert \psi_0 \rangle = \sin \left( 2\theta + \frac{\pi}{4} \right) \vert V \rangle_{\sigma} + \cos \left( 2\theta + \frac{\pi}{4} \right) \vert H \rangle_{\sigma} \\
\vert \psi_1 \rangle = \sin \left( 2\theta - \frac{\pi}{4} \right) \ket{V}_{\sigma} - \cos \left( 2\theta - \frac{\pi}{4} \right) \ket{H}_{\sigma} 
\end{aligned}
\end{equation}
are the states of the signal after the meter measurement. Similar expressions hold for the application of $N_k$. Such states are useful for the calculation of the lower bound on the irreversible entropy, which requires the evaluation of the residual entropy associated to the post-measurement state of the signal. Such a lower bound does not depend on the von Neumann entropy of the signal state as $\rho_{\sigma}$ is pure. In the limiting case of $T\rightarrow 0$, the residual entropy reduces to the opposite of the Shannon entropy, thus giving a null GO (QC) information. As the temperature of the thermal bath rises, the correlation term accounts also for the mixedness of the state of the meter.

The quantity in the left-hand side of Eq.~\eqref{SU}  can be interpreted as an irreversible entropy $S_{irr}=\beta(W_\mu^{meas}-\Delta F_{\mu})$. We have $
W_{\mu}^{meas}=\sum_k p_k \Tr[\rho_k^{\mu} \mathbb{H}_{\mu}^k] - \Tr[\rho_{\mu}^{init} \mathbb{H}_{\mu}]$
where, $\rho_k^{\mu}=\ket{k}\bra{k}$, $\mathbb{H}_{\mu}^k=\epsilon_k \ket{k}\bra{k}$, and $\rho_{\mu}^{init}=e^{-\beta \mathbb{H}_{\mu}}/\mathbb{Z}_{\mu}$. The average change in the Helmholtz free energy is $\Delta F_{\mu}=\sum_k p_k F_{\mu}^k
- F_{\mu}^{init}$, in which $F_{\mu}^k=-k_B T \ln \mathbb{Z}_{\mu}^k$. These lead to the following expressions
\begin{equation}
\begin{aligned}
&W_{\mu}^{meas}=p_1 \bar{\Delta} + \epsilon_0 - \langle \epsilon \rangle,\\
&\Delta F_{\mu}=p_1 \bar{\Delta} + \beta^{-1} ln (1+e^{-\beta \bar{\Delta}}),
\end{aligned}
\end{equation}
where $\bar{\Delta}=\epsilon_1-\epsilon_0$ and $\langle \epsilon \rangle=\Tr[\rho_{\mu}^{init} \mathbb{H}_{\mu}]$. Remarkably, in line with the lower bound in Eq.~\eqref{SU}, the irreversible entropy does not depend on the invasiveness of the measurement process. In fact, we get the expression
\begin{equation}
S_{irr}=\beta \epsilon_0 - \beta \langle \epsilon \rangle - ln(1+e^{-\beta \bar{\Delta}}).
\end{equation}


\textit{Methods: experimental details.} We built our investigation on the possibility to use pair of photons to encode both signal and meter systems. Photons are generated via a Spontaneous Parametric Down Conversion (SPDC) nonlinear process through a $\beta$ Barium-Borate crystal (Type I): the source produces photon pairs at 810 nm when pumped with a 405 nm Continuous-Wave (CW) laser at 80 mW. The two photons interact in a C-Sign gate realized using Partially Polarized Beam Splitters (PPBSs) with horizontal (vertical) polarization trasmittivity $T_H (T_V)=1(1/3)$ and Hadamard gates. This experimental gate acts transforming $\vert HH \rangle \rightarrow 1/3 \vert VV \rangle$, $\vert HV \rangle \rightarrow 1/3 \vert VH \rangle$, $\vert VH \rangle \rightarrow 1/3 \vert HV \rangle$, $\vert VV \rangle \rightarrow -1/3\vert HH \rangle$. Upon suitable normalization of the detected coincidence counts, we have the following probabilities
\begin{equation}
p_D=\left( N_{HD}+N_{VD} \right)/N_0 \;\;\; p_A=\left( N_{HA}+N_{VA} \right)/N_0
\end{equation}
with $N_0=N_{HD}+N_{VD}+N_{HA}+N_{VA}$. These are instrumental to the evaluation of the Shannon entropies. Performing a Quantum State Tomography (QST) procedure on the signal qubit after the measurement of the meter allows to obtain the experimental version of the $\tilde{\rho}_{\sigma}^k$ density matrix in Eq. \ref{Revisited}, giving us direct access to the residual entropy of the post-measurement states.


\bigskip


 
{\it Acknowledgments.} Mauro Paternostro is supported by the EU Collaborative Project TherMiQ (Grant Agreement 618074), the Julian Schwinger Foundation (grant Nr. JSF-14-7-0000), the Royal Society Newton Mobility Grant (grant NI160057), the DfE-SFI Investigator Programme (grant 15/IA/2864). Part of this work was supported by COST Action MP1209 Thermodynamics in the quantum regime. Marco Barbieri is supported by a Rita Levi-Montalcini fellowship of MIUR. We thank Roberto Raimondi, Maria Antonietta Ricci, and Fabio Bruni for useful discussions and comments, Paolo Aloe for technical assistance, Carlo Meneghini and Francesca Paolucci for the loan of scientific equipment.

\end{document}